\documentclass{article}

\usepackage{graphicx}
\usepackage[round]{natbib}
\usepackage{epsfig}

\title{
Objective Climate Model Predictions Using Jeffreys' Prior:
the General Multivariate Normal Case
}

\author{Stephen Jewson\footnote{\emph{Correspondence email}: \texttt{stephen.jewson@rms.com}},
Dan Rowlands, Myles Allen\\}

\begin{document}
\maketitle

\begin{abstract}
Objective probabilistic forecasts of future climate that include parameter uncertainty
can be made by using the Bayesian prediction integral with the prior set to Jeffreys' Prior.
The calculations involved in determining the prior can then be simplified by making parametric assumptions about the distribution
of the output from the climate model. The most obvious assumption to make is that the climate model
output is normally distributed, in which case evaluating the prior becomes a question of evaluating
gradients in the parameters of the normal distribution. In previous work we have considered the
special cases of
diagonal (but not constant) covariance matrix, and constant (but not diagonal) covariance matrix.
We now derive expressions for the general multivariate normal distribution,
with non-constant non-diagonal covariance matrix. The algebraic manipulation required is
more complex than for the special cases, and involves some slightly esoteric matrix operations including
taking the expectation of a vector quadratic form and differentiating the determinants, traces and inverses of matrices.
\end{abstract}

\section{Introduction}

We are interested in using climate models to produce objective probabilistic forecasts of future climate that
include parameter uncertainty. The word `objective' is used here in a technical
statistical sense that means that the prior distribution for the parameters is determined by a rule,
rather than from intuition.
In statistics the most widely discussed rule is the Jeffreys' rule~\citep{jeffreys}.
Jeffreys' rule could, in principal, be applied to a climate model directly from the definition.
This would translate into using numerical methods to differentiate the predicted probabilities from
initial condition ensembles and to take expectations over all simulated climate states.
That could, however, be computationally demanding.
As an alternative one could fit distributions to the output from the model, and differentiate the estimated parameters of
the distributions instead, which is likely to be considerably easier.
The most obvious distribution to fit is then the multivariate normal distribution.
In previous work we have considered two special cases.
In~\citet{jp1} we considered the case where the predicted variables are independent, but both the mean and the variance of initial condition
ensembles from the model are allowed to vary as a function of the model parameters.
In~\citet{jp2} we considered the complementary case where the predicted variables can be correlated, but with a covariance matrix that is constant
as a function of the model parameters.
We now consider the general case, with correlated predicted variables \emph{and} a covariance matrix that can vary as a function of the
model parameters.
This general case contains the two special cases.
The algebra in this case is not quite as elementary as the two special cases, in that we have
to take expectations of a vector quadratic form, and differentiate the determinants, traces and inverses of matrices.
Pedagogically, therefore, we consider the derivations used in the two simpler cases as still being useful, especially
as climate modelling practice is perhaps unlikely to reach the stage where it would be necessary to consider the covariance
matrix varying as a function of parameters for quite some time (there are many other more important challenges to be dealt with first).

Detailed explanations of Jeffreys' Prior, and the motivation for its use, are given in~\citet{jp1} and will not be repeated here.
In section~\ref{s2} below we give the expressions for Jeffreys' Prior and the multivariate normal density.
In section~\ref{s3} we then derive the expression for Jeffreys' Prior for the multivariate normal distribution where the climate model
has just a single parameter.
In section~\ref{s4} we derive the same expression, but considering multiple parameters.
In both section~\ref{s3} and section~\ref{s4} we also consider four special cases:
independence (taking us back to the results in~\citet{jp1}),
constant covariance (taking us back to the results in~\citet{jp2}),
constant correlation and
constant variance.

\section{Jeffreys' Prior and the Multivariate Normal Density}\label{s2}

We start with definitions of the Jeffreys' Prior and the multivariate normal density.

\subsection{Jeffreys' Prior}

Jeffreys' Prior is given by:

\begin{equation}\label{eq1}
    p(\theta)=\sqrt{\mbox{det}\left[-\mbox{E}\left(\frac{\partial^2 \log p(x|\theta)}{\partial \theta_j \partial \theta_k}\right)\right]}\\
    \label{jp_def}
\end{equation}

Detailed explanations of this equation, which can at first be rather difficult to understand, are given in~\citet{jp1} and~\citet{jp2},
as well as many statistics textbooks such as~\citet{peterlee}.

\subsection{The Multivariate Normal Density}

Probability densities from the multivariate normal distribution are given by:
\begin{equation}
p(x|\theta)=\frac{1}{(2\pi)^\frac{n}{2}} \frac{1}{D^{\frac{1}{2}}} \mbox{exp}\left(-\frac{1}{2}(x-\mu)^T S(x-\mu)\right)
\end{equation}

where, in our application:
\begin{itemize}
  \item $x$ is a vector of those variables predicted by the climate model that are to be compared
  with observations
  \item $\theta$ is a vector for the underlying parameters in the climate model
  \item $\mu$=$\mu(\theta)$ is a vector of the mean response of the model (in other words, the ensemble mean
  of $x$ for an infinite-sized initial condition ensemble for fixed parameters $\theta$)
  \item $\Sigma=\Sigma(\theta)$ is the covariance matrix of the response of the model
   (in other words, the ensemble covariance matrix of $x$ for an infinite-sized initial condition ensemble for fixed parameters $\theta$)
  \item $S=\Sigma^{-1}=S(\theta)$ is the inverse of the covariance matrix
  \item $D=\mbox{det}(\Sigma)=D(\theta)$ is the determinant of the covariance matrix
\end{itemize}

This gives:
\begin{eqnarray}
\ln p(x|\theta)
&=&-\frac{n}{2}\ln 2\pi -\frac{1}{2} \ln D -\frac{1}{2}(x-\mu)^T S(x-\mu)\\
&=&-\frac{n}{2}\ln 2\pi -\frac{1}{2} \ln D -\frac{1}{2}(x^T S x-x^TS \mu-\mu^T S x+\mu^T S \mu)
\end{eqnarray}

Since $S$ is symmetric we have $x^TS\mu=\mu^T S x$
which means that the above expression for $\ln p(x|\theta)$ simplifies a little to:
\begin{eqnarray}
\ln p(x|\theta)\label{eq1}
&=&-\frac{n}{2}\ln 2\pi -\frac{1}{2} \ln D -\frac{1}{2}(x^T S x-2x^TS \mu+\mu^T S \mu)
\end{eqnarray}

We now consider two cases: one parameter and multiple parameters.

\section{One parameter}\label{s3}

The first case we consider is where there is just a single parameter in the climate model.
We mainly consider this case as a warm-up for the multiple parameter case, although it would also be
relevant if one only wanted to model the uncertainty due to a single parameter, which might be
a good approximation to the overall parameter uncertainty if that single parameter dominates the uncertainty.

If we consider $\theta$ to be this single (scalar) parameter, then differentiating equation~\ref{eq1} by $\theta$ gives:
\begin{eqnarray}
\frac{\partial \ln p(x|\theta)}{\partial \theta}
&=&-\frac{1}{2} \frac{\partial \ln D}{\partial \theta}
   -\frac{1}{2}x^T \frac{\partial S}{\partial \theta} x
   +x^T\frac{\partial (S \mu)}{\partial \theta}
   -\frac{1}{2}\frac{\partial (\mu^T S \mu)}{\partial \theta}
\end{eqnarray}

Expanding the third and fourth terms:
\begin{eqnarray}
\frac{\partial \ln p(x|\theta)}{\partial \theta}
&=&-\frac{1}{2} \frac{\partial \ln D}{\partial \theta}
   -\frac{1}{2}x^T \frac{\partial S}{\partial \theta} x
   +x^T \left(\frac{\partial S}{\partial \theta}\mu+S\frac{\partial \mu}{\partial \theta}\right)
   -\frac{1}{2}\left(\frac{\partial \mu^T }{\partial \theta}S \mu+\mu^T\frac{\partial S }{\partial \theta}\mu+\mu^T S \frac{\partial \mu}{\partial \theta}\right)\\
&=&-\frac{1}{2} \frac{\partial \ln D}{\partial \theta}
   -\frac{1}{2}x^T \frac{\partial S}{\partial \theta} x
   +x^T \frac{\partial S}{\partial \theta}\mu+x^T S\frac{\partial \mu}{\partial \theta}
   -\frac{1}{2}\frac{\partial \mu^T }{\partial \theta}S \mu-\frac{1}{2}\mu^T\frac{\partial S }{\partial \theta}\mu-\frac{1}{2}\mu^T S \frac{\partial \mu}{\partial \theta}
\end{eqnarray}

Since $S$ is symmetric $\mu^T S \frac{\partial \mu}{\partial \theta}=\frac{\partial \mu^T }{\partial \theta}S \mu$ and so
the 5th and 7th terms combine, giving:

\begin{eqnarray}\label{eq10}
\frac{\partial \ln p(x|\theta)}{\partial \theta}
&=&-\frac{1}{2} \frac{\partial \ln D}{\partial \theta}
   -\frac{1}{2}x^T \frac{\partial S}{\partial \theta} x
   +x^T \frac{\partial S}{\partial \theta}\mu
   +x^T S\frac{\partial \mu}{\partial \theta}
   -\frac{\partial \mu^T }{\partial \theta}S \mu
   -\frac{1}{2}\mu^T\frac{\partial S }{\partial \theta}\mu
\end{eqnarray}

Differentiating again wrt $\theta$ gives:
\begin{eqnarray}
\frac{\partial^2 \ln p(x|\theta)}{\partial \theta^2}
&=&-\frac{1}{2} \frac{\partial^2 \ln D}{\partial \theta^2}
   -\frac{1}{2}x^T \frac{\partial^2 S}{\partial \theta^2} x
   +x^T \frac{\partial}{\partial \theta}\left(\frac{\partial S}{\partial \theta}\mu\right)
   +x^T \frac{\partial}{\partial \theta}\left(S\frac{\partial \mu}{\partial \theta}\right)\\
&& -\frac{\partial}{\partial \theta}\left(\frac{\partial \mu^T }{\partial \theta}S \mu\right)
   -\frac{1}{2}\frac{\partial}{\partial \theta}\left(\mu^T\frac{\partial S }{\partial \theta}\mu\right)\\
&=&-\frac{1}{2} \frac{\partial^2 \ln D}{\partial \theta^2}
   -\frac{1}{2}x^T \frac{\partial^2 S}{\partial \theta^2} x
   +x^T \left(\frac{\partial^2 S}{\partial \theta^2}\mu+\frac{\partial S}{\partial \theta}\frac{\partial \mu}{\partial \theta}\right)
   +x^T \left(\frac{\partial S}{\partial \theta}\frac{\partial \mu}{\partial \theta}+S\frac{\partial^2 \mu}{\partial \theta^2}\right)\\
&&   -\left(\frac{\partial^2 \mu^T }{\partial \theta^2}S \mu
                    +\frac{\partial \mu^T }{\partial \theta}\frac{\partial S}{\partial \theta} \mu
                    +\frac{\partial \mu^T }{\partial \theta}S \frac{\partial \mu}{\partial \theta}\right)
   -\frac{1}{2}\left(\frac{\partial \mu^T}{\partial \theta}\frac{\partial S }{\partial \theta}\mu
         +\mu^T\frac{\partial^2 S }{\partial \theta^2}\mu
         +\mu^T\frac{\partial S }{\partial \theta}\frac{\partial \mu}{\partial \theta}\right)\\
&&\mbox{(where we have expanded the derivatives of products)}\nonumber\\
&=&-\frac{1}{2} \frac{\partial^2 \ln D}{\partial \theta^2}
   -\frac{1}{2}x^T \frac{\partial^2 S}{\partial \theta^2} x
   +x^T \frac{\partial^2 S}{\partial \theta^2}\mu
   +x^T \frac{\partial S}{\partial \theta}\frac{\partial \mu}{\partial \theta}
   +x^T \frac{\partial S}{\partial \theta}\frac{\partial \mu}{\partial \theta}
   +x^T S\frac{\partial^2 \mu}{\partial \theta^2}\\
&&   -\frac{\partial^2 \mu^T }{\partial \theta^2}S \mu
     -\frac{\partial \mu^T }{\partial \theta}\frac{\partial S}{\partial \theta} \mu
     -\frac{\partial \mu^T }{\partial \theta}S \frac{\partial \mu}{\partial \theta}
   -\frac{1}{2}\frac{\partial \mu^T}{\partial \theta}\frac{\partial S }{\partial \theta}\mu
   -\frac{1}{2}\mu^T\frac{\partial^2 S }{\partial \theta^2}\mu
   -\frac{1}{2}\mu^T\frac{\partial S }{\partial \theta}\frac{\partial \mu}{\partial \theta}\\
&=&-\frac{1}{2} \frac{\partial^2 \ln D}{\partial \theta^2}
   -\frac{1}{2}x^T \frac{\partial^2 S}{\partial \theta^2} x
   +x^T \frac{\partial^2 S}{\partial \theta^2}\mu
   +2x^T \frac{\partial S}{\partial \theta}\frac{\partial \mu}{\partial \theta}
   +x^T S\frac{\partial^2 \mu}{\partial \theta^2}\\
&&   -\frac{\partial^2 \mu^T }{\partial \theta^2}S \mu
     -2\frac{\partial \mu^T }{\partial \theta}\frac{\partial S}{\partial \theta} \mu
     -\frac{\partial \mu^T }{\partial \theta}S \frac{\partial \mu}{\partial \theta}
     -\frac{1}{2}\mu^T\frac{\partial^2 S }{\partial \theta^2}\mu\\
&&\mbox{(where we have combined the 4th and 5th, and 8th, 10th and 12th terms)}\nonumber
\end{eqnarray}

Taking expectations:
\begin{eqnarray}
E\left(\frac{\partial^2 \ln p(x|\theta)}{\partial \theta^2}\right)
&=&-\frac{1}{2} \frac{\partial^2 \ln D}{\partial \theta^2}
   -\frac{1}{2}E\left(x^T \frac{\partial^2 S}{\partial \theta^2} x\right)
   +E(x^T) \frac{\partial^2 S}{\partial \theta^2}\mu
   +2E(x^T) \frac{\partial S}{\partial \theta}\frac{\partial \mu}{\partial \theta}\\
&&   +E(x^T) S\frac{\partial^2 \mu}{\partial \theta^2}
    -\frac{\partial^2 \mu^T }{\partial \theta^2}S \mu
     -2\frac{\partial \mu^T }{\partial \theta}\frac{\partial S}{\partial \theta} \mu
     -\frac{\partial \mu^T }{\partial \theta}S \frac{\partial \mu}{\partial \theta}
   -\frac{1}{2}\mu^T\frac{\partial^2 S }{\partial \theta^2}\mu
\end{eqnarray}

But $E(x)=\mu$, and, from appendix 2,
\begin{equation}
E\left(x^T \frac{\partial^2 S}{\partial \theta^2} x\right)=\mbox{tr}\left( \frac{\partial^2 S}{\partial \theta^2}\Sigma\right)+\mu^T \frac{\partial^2 S}{\partial \theta^2} \mu
\end{equation}

and so:

\begin{eqnarray}
E\left(\frac{\partial^2 \ln p(x|\theta)}{\partial \theta^2}\right)
&=&-\frac{1}{2} \frac{\partial^2 \ln D}{\partial \theta^2}
   -\frac{1}{2} \left(\mbox{tr}\left(\frac{\partial^2 S}{\partial \theta^2}\Sigma \right)+\mu^T \frac{\partial^2 S}{\partial \theta^2} \mu\right)
   +\mu^T \frac{\partial^2 S}{\partial \theta^2}\mu
   +2\mu^T \frac{\partial S}{\partial \theta}\frac{\partial \mu}{\partial \theta}\\
&&   +\mu^T S\frac{\partial^2 \mu}{\partial \theta^2}
   -\frac{\partial^2 \mu^T }{\partial \theta^2}S \mu
     -2\frac{\partial \mu^T }{\partial \theta}\frac{\partial S}{\partial \theta} \mu
     -\frac{\partial \mu^T }{\partial \theta}S \frac{\partial \mu}{\partial \theta}
   -\frac{1}{2}\mu^T\frac{\partial^2 S }{\partial \theta^2}\mu\\
&=&-\frac{1}{2} \frac{\partial^2 \ln D}{\partial \theta^2}
   -\frac{1}{2} \mbox{tr}\left( \frac{\partial^2 S}{\partial \theta^2}\Sigma\right)
   -\frac{1}{2} \mu^T \frac{\partial^2 S}{\partial \theta^2} \mu
   +\mu^T \frac{\partial^2 S}{\partial \theta^2}\mu
   +2\mu^T \frac{\partial S}{\partial \theta}\frac{\partial \mu}{\partial \theta}\\
&&   +\mu^T S\frac{\partial^2 \mu}{\partial \theta^2}
   -\frac{\partial^2 \mu^T }{\partial \theta^2}S \mu
     -2\frac{\partial \mu^T }{\partial \theta}\frac{\partial S}{\partial \theta} \mu
     -\frac{\partial \mu^T }{\partial \theta}S \frac{\partial \mu}{\partial \theta}
   -\frac{1}{2}\mu^T\frac{\partial^2 S }{\partial \theta^2}\mu\\\label{eq2}
&=&-\frac{1}{2} \left[\frac{\partial^2 \ln D}{\partial \theta^2}+\mbox{tr}\left( \frac{\partial^2 S}{\partial \theta^2}\Sigma\right)\right]
   -\frac{\partial \mu^T }{\partial \theta}S \frac{\partial \mu}{\partial \theta}\\
&&\mbox{(where we have cancelled various terms)}\nonumber
\end{eqnarray}

But
\begin{eqnarray}
\frac{\partial^2 \ln D}{\partial \theta^2}
&=&\frac{\partial}{\partial \theta}\left(\frac{\partial \log D}{\partial \theta}\right)\\
&=&\frac{\partial}{\partial \theta}\left(\frac{1}{D}\frac{\partial D}{\partial \theta}\right)\\
&=&\frac{\partial}{\partial \theta}\left(\frac{1}{D}\left(D\mbox{tr}\left(S\frac{\partial \Sigma}{\partial \theta}\right)\right)\right)\\
&&\mbox{(using a standard result for the derivative of a determinant known as Jacobi's formula)}\nonumber\\
&=&\frac{\partial}{\partial \theta}\left(\mbox{tr}\left(S\frac{\partial \Sigma}{\partial \theta}\right)\right)\\
&=&\mbox{tr}\left(\frac{\partial}{\partial \theta}\left(S\frac{\partial \Sigma}{\partial \theta}\right)\right)\\
&&\mbox{(using a standard result for the derivative of a trace)}\nonumber\\
&=&\mbox{tr}\left(\frac{\partial S}{\partial \theta}\frac{\partial \Sigma}{\partial \theta}+S\frac{\partial^2 \Sigma}{\partial \theta^2}\right)\\
&=&\mbox{tr}\left(\frac{\partial S}{\partial \theta}\frac{\partial \Sigma}{\partial \theta}\right)+\mbox{tr}\left( S \frac{\partial^2 \Sigma}{\partial \theta^2}\right)\\
&&\mbox{(using the linearity of the trace operator)}\nonumber
\end{eqnarray}

Now note that
\begin{eqnarray}
\frac{\partial}{\partial \theta}(S \Sigma)&=&S \frac{\partial \Sigma}{\partial \theta}+\frac{\partial S}{\partial \theta}\Sigma
\end{eqnarray}
and
\begin{eqnarray}
\frac{\partial^2}{\partial \theta^2}( S \Sigma)
&=&S \frac{\partial^2 \Sigma}{\partial \theta^2}+\frac{\partial^2 S}{\partial \theta^2}\Sigma+2\frac{\partial S}{\partial \theta}\frac{\partial \Sigma}{\partial \theta}
\end{eqnarray}

But $S\Sigma=I$ and so $\frac{\partial^2}{\partial \theta^2}( S \Sigma)=0$, implying that
\begin{equation}
S\frac{\partial^2 \Sigma}{\partial \theta^2}=-\frac{\partial^2 S}{\partial \theta^2}\Sigma -2\frac{\partial S}{\partial \theta}\frac{\partial \Sigma}{\partial \theta}
\end{equation}

and so
\begin{eqnarray}
\frac{\partial^2 \ln D}{\partial \theta^2}
&=&\mbox{tr}\left(\frac{\partial S}{\partial \theta}\frac{\partial \Sigma}{\partial \theta}\right)+\mbox{tr}\left( S \frac{\partial^2 \Sigma}{\partial \theta^2}\right)\\
&=&\mbox{tr}\left(\frac{\partial S}{\partial \theta}\frac{\partial \Sigma}{\partial \theta}\right)+\mbox{tr}\left(- \frac{\partial^2 S}{\partial \theta^2}-2\frac{\partial S}{\partial \theta}\frac{\partial \Sigma}{\partial \theta}\Sigma\right)\\
&=&\mbox{tr}\left(\frac{\partial S}{\partial \theta}\frac{\partial \Sigma}{\partial \theta}\right)-\mbox{tr}\left( \frac{\partial^2 S}{\partial \theta^2}\Sigma\right)-\mbox{tr}\left(2\frac{\partial S}{\partial \theta}\frac{\partial \Sigma}{\partial \theta}\right)\\
&=&-\mbox{tr}\left(\frac{\partial^2 S}{\partial \theta^2}\Sigma \right)-\mbox{tr}\left(\frac{\partial S}{\partial \theta}\frac{\partial \Sigma}{\partial \theta}\right)
\end{eqnarray}

Giving:
\begin{eqnarray}
\frac{\partial^2 \ln D}{\partial \theta^2}+\mbox{tr}\left( \frac{\partial^2 S}{\partial \theta^2}\Sigma\right)
&=&\mbox{tr}\left(- \frac{\partial^2 S}{\partial \theta^2}\Sigma\right)-\mbox{tr}\left(\frac{\partial S}{\partial \theta}\frac{\partial \Sigma}{\partial \theta}\right)+\mbox{tr}\left(S \frac{\partial^2 \Sigma}{\partial \theta^2}\right)\\\label{eq3}
&=&-\mbox{tr}\left(\frac{\partial S}{\partial \theta}\frac{\partial \Sigma}{\partial \theta}\right)
\end{eqnarray}

Returning to equation~\ref{eq2} and substituting in the expression given in equation~\ref{eq3} gives:

\begin{eqnarray}
E\left(\frac{\partial^2 \ln p(x|\theta)}{\partial \theta^2}\right)
&=&-\frac{1}{2} \left[\frac{\partial^2 \ln D}{\partial \theta^2}+\mbox{tr}\left( \frac{\partial^2 S}{\partial \theta^2}\Sigma\right)\right]
   -\frac{\partial \mu^T }{\partial \theta}S \frac{\partial \mu}{\partial \theta}\\
&=&-\frac{1}{2} \left[-\mbox{tr}\left(\frac{\partial S}{\partial \theta}\frac{\partial \Sigma}{\partial \theta}\right)\right]
   -\frac{\partial \mu^T }{\partial \theta}S \frac{\partial \mu}{\partial \theta}\\
&=&\frac{1}{2}\mbox{tr}\left(\frac{\partial S}{\partial \theta}\frac{\partial \Sigma}{\partial \theta}\right)
   -\frac{\partial \mu^T }{\partial \theta}S \frac{\partial \mu}{\partial \theta}\\
&=&-\frac{1}{2}\mbox{tr}\left(S\frac{\partial \Sigma}{\partial \theta}S\frac{\partial \Sigma}{\partial \theta}\right)
   -\frac{\partial \mu^T }{\partial \theta}S \frac{\partial \mu}{\partial \theta}\\
   &&\mbox{(using the standard result that $\frac{\partial A^{-1}}{\partial \theta}=-A^{-1}\frac{\partial A}{\partial \theta}A^{-1}$)}
\end{eqnarray}

This gives the prior:
\begin{eqnarray}
p(\theta)
&=&\sqrt{-E\left(\frac{\partial^2 \ln p(x|\theta)}{\partial \theta^2}\right)}\\
&=&\sqrt{\frac{\partial \mu^T }{\partial \theta}S \frac{\partial \mu}{\partial \theta}
-\frac{1}{2}\mbox{tr}\left(\frac{\partial S}{\partial \theta}\frac{\partial \Sigma}{\partial \theta}\right)}\label{eq100}\\
&=&\sqrt{\frac{\partial \mu^T }{\partial \theta}S \frac{\partial \mu}{\partial \theta}
+\frac{1}{2}\mbox{tr}\left(S\frac{\partial \Sigma}{\partial \theta}S\frac{\partial \Sigma}{\partial \theta}\right)}\label{eq100b}
\end{eqnarray}

If there are $n$ observations then:
\begin{itemize}
  \item $\mu$ is an $n$ x $1$ vector
  \item $\frac{\partial \mu}{\partial \theta}$ is an $n$ by $1$ vector
  \item $\mu^T$ is a $1$ x $n$ vector
  \item $\frac{\partial \mu^T}{\partial \theta}$ is a $1$ by $n$ vector
  \item $S$ is an $n$ by $n$ matrix
  \item $\frac{\partial \mu^T }{\partial \theta}S \frac{\partial \mu}{\partial \theta}$ is a scalar
  \item $\frac{\partial S}{\partial \theta}$ is an $n$ by $n$ matrix
  \item $\frac{\partial \Sigma}{\partial \theta}$ is an $n$ by $n$ matrix
  \item $\frac{\partial S}{\partial \theta}\frac{\partial \Sigma}{\partial \theta}$ is an $n$ by $n$ matrix
  \item tr$\left(\frac{\partial S}{\partial \theta}\frac{\partial \Sigma}{\partial \theta}\right)$ is a scalar
\end{itemize}

We now consider various special cases of this general formula, starting with the two cases described in~\citet{jp1} and~\citet{jp2}.

\subsection{Independence}

If the observations are modelled as independent then:
\begin{eqnarray}
\Sigma
&=&\mbox{diag}\left(\sigma_1^2,\sigma_2^2,...,\sigma_n^2\right)\\
S
&=&\mbox{diag}\left(\sigma_1^{-2},\sigma_2^{-2},...,\sigma_n^{-2}\right)\\
\frac{\partial \mu^T }{\partial \theta}S \frac{\partial \mu}{\partial \theta}
&=&\sum_{i=1}^{n}\frac{1}{\sigma_i^2}\left(\frac{\partial \mu_i}{\partial \theta}\right)^2\\
\frac{\partial \Sigma}{\partial \theta}
&=&\mbox{diag}\left(2\sigma_1 \frac{\partial \sigma_1}{\partial \theta},2\sigma_2\frac{\partial \sigma_2}{\partial \theta},...,2\sigma_n\frac{\partial \sigma_n}{\partial \theta}\right)\\
\frac{\partial S}{\partial \theta}
&=&\mbox{diag}\left(-2\sigma_1^{-3} \frac{\partial \sigma_1}{\partial \theta},-2\sigma_2^{-3}\frac{\partial \sigma_2}{\partial \theta},...,-2\sigma_n^{-3}\frac{\partial \sigma_n}{\partial \theta}\right)\\
\frac{\partial S}{\partial \theta}\frac{\partial \Sigma}{\partial \theta}
&=&\mbox{diag}\left(-4\sigma_1^{-2} \frac{\partial \sigma_1}{\partial \theta},-4\sigma_2^{-2}\frac{\partial \sigma_2}{\partial \theta},...,-4\sigma_n^{-2}\frac{\partial \sigma_n}{\partial \theta}\right)\\
&=&-4\mbox{diag}\left(\frac{1}{\sigma_1^2} \frac{\partial \sigma_1}{\partial \theta},\frac{1}{\sigma_2^2}\frac{\partial \sigma_2}{\partial \theta},...,\frac{1}{\sigma_n^2}\frac{\partial \sigma_n}{\partial \theta}\right)\\
-\frac{1}{2}\mbox{tr}\left(\frac{\partial S}{\partial \theta}\frac{\partial \Sigma}{\partial \theta}\right)
&=&2\sum_{i=1}^{n}\frac{1}{\sigma_i^2}\left(\frac{\partial \sigma_i}{\partial \theta}\right)^2
\end{eqnarray}

and so the prior is
\begin{eqnarray}
p(\theta)&=&\sqrt{\sum_{i=1}^{n}\frac{1}{\sigma_i^2}\left(\frac{\partial \mu_i}{\partial \theta}\right)^2
               +2\sum_{i=1}^{n}\frac{1}{\sigma_i^2}\left(\frac{\partial \sigma_i}{\partial \theta}\right)^2}
\end{eqnarray}

which agrees with equation 19 in~\citet{jp1}.

\subsection{Constant Covariance}

If the covariance $\Sigma$ is constant then equation~\ref{eq100} reduces immediately to
\begin{eqnarray}
p(\theta)&=&\sqrt{\frac{\partial \mu^T }{\partial \theta}S \frac{\partial \mu}{\partial \theta}}
\end{eqnarray}

which agrees with equation 19 in~\citet{jp2}.

\subsection{Constant Correlation}

If the correlations are modelled as constant (but the variances are allowed to vary) then:

\begin{eqnarray}
\Sigma&=&VCV\\
\mbox{where }V&=&\mbox{diag}(\sigma_1^2,\sigma_2^2,...,\sigma_n^2)\\
\mbox{and }C&=&\mbox{the correlation matrix}\\
\frac{\partial \Sigma}{\partial \theta}
&=& \frac{\partial V}{\partial \theta}CV+VC\frac{\partial V}{\partial \theta}\\
&=& 2\frac{\partial V}{\partial \theta}CV\\
\mbox{where } \frac{\partial V}{\partial \theta}
&=&2\mbox{diag}(\sigma_1\frac{\partial \sigma_1}{\partial \theta},\sigma_2\frac{\partial \sigma_2}{\partial \theta},...,\sigma_n\frac{\partial \sigma_n}{\partial \theta})\\
S&=&V^{-1} C^{-1} V^{-1}\\
\mbox{where }V^{-1}&=&\mbox{diag}(\sigma_1^{-2},\sigma_2^{-2},...,\sigma_n^{-2})\\
\mbox{and }C^{-1}&=&\mbox{the inverse correlation matrix}\\
\frac{\partial \Sigma}{\partial \theta}
&=& \frac{\partial V^{-1}}{\partial \theta}C^{-1}V^{-1}+V^{-1}C^{-1}\frac{\partial V^{-1}}{\partial \theta}\\
&=& 2V^{-1}C^{-1}\frac{\partial V^{-1}}{\partial \theta}\\
\mbox{where } \frac{\partial V^{-1}}{\partial \theta}
&=&-2\mbox{diag}\left(\frac{1}{\sigma_1^2}\frac{\partial \sigma_1}{\partial \theta},\frac{1}{\sigma_2^2}\frac{\partial \sigma_2}{\partial \theta},...,\frac{1}{\sigma_n^2}\frac{\partial \sigma_n}{\partial \theta}\right)\\
\frac{\partial S}{\partial \theta}\frac{\partial \Sigma}{\partial \theta}
&=&4 V^{-1}C^{-1}\frac{\partial V^{-1}}{\partial \theta}\frac{\partial V}{\partial \theta}CV\\
\frac{\partial V^{-1}}{\partial \theta}\frac{\partial V}{\partial \theta}
&=&-4\mbox{diag}\left(\frac{1}{\sigma_1}\left(\frac{\partial \sigma_1}{\partial \theta}\right)^2,\frac{1}{\sigma_2}\left(\frac{\partial \sigma_2}{\partial \theta}\right)^2,...,\frac{1}{\sigma_n}\left(\frac{\partial \sigma_n}{\partial \theta}\right)^2\right)
\end{eqnarray}

and the prior is:
\begin{eqnarray}
p(\theta)
&=&\sqrt{\frac{\partial \mu^T }{\partial \theta}S \frac{\partial \mu}{\partial \theta}
-\frac{1}{2}\mbox{tr}\left(4 V^{-1}C^{-1}\frac{\partial V^{-1}}{\partial \theta}\frac{\partial V}{\partial \theta}CV\right)}
\end{eqnarray}

\subsection{Constant Variance}

If the variances are modelled as constant (but the correlations are allowed to vary) then:
\begin{eqnarray}
\Sigma&=&VCV\\
\mbox{where }V&=&\mbox{diag}(\sigma_1^2,\sigma_2^2,...,\sigma_n^2)\\
\mbox{and }C&=&\mbox{the correlation matrix}\\
\frac{\partial \Sigma}{\partial \theta}&=& V \frac{\partial C}{\partial \theta}V\\
S&=&V^{-1} C^{-1} V^{-1}\\
\mbox{where }V^{-1}&=&\mbox{diag}(\sigma_1^{-2},\sigma_2^{-2},...,\sigma_n^{-2})\\
\mbox{and }C^{-1}&=&\mbox{the inverse correlation matrix}\\
\frac{\partial S}{\partial \theta}&=&V^{-1} \frac{\partial C^{-1}}{\partial \theta}V^{-1}\\
\frac{\partial S}{\partial \theta}\frac{\partial \Sigma}{\partial \theta}
&=&V^{-1}\frac{\partial C^{-1}}{\partial \theta}V^{-1}V\frac{\partial C}{\partial \theta}V\\
&=&V^{-1}\frac{\partial C^{-1}}{\partial \theta}\frac{\partial C}{\partial \theta}V\\
&=&-V^{-1}C^{-1}\frac{\partial C}{\partial \theta}C^{-1}\frac{\partial C}{\partial \theta}V
\end{eqnarray}

and so the prior is:
\begin{eqnarray}
p(\theta)
&=&\sqrt{\frac{\partial \mu^T }{\partial \theta}S \frac{\partial \mu}{\partial \theta}
-\frac{1}{2}\mbox{tr}\left(V^{-1}\frac{\partial C^{-1}}{\partial \theta}\frac{\partial C}{\partial \theta}V\right)}\label{eq101}\\
&=&\sqrt{\frac{\partial \mu^T }{\partial \theta}S \frac{\partial \mu}{\partial \theta}
+\frac{1}{2}\mbox{tr}\left(V^{-1}C^{-1}\frac{\partial C}{\partial \theta}C^{-1}\frac{\partial C}{\partial \theta}V\right)}\label{eq101b}
\end{eqnarray}

\section{Multiple Parameters}\label{s4}

We now consider the multiparameter case.
We start with two parameters and generalise to multiple parameters later.
The derivations are only slightly more complex than those for the single parameter case.
Starting from equation~\ref{eq10}, which was:

\begin{eqnarray}
\frac{\partial \ln p(x|\theta)}{\partial \theta}
&=&-\frac{1}{2} \frac{\partial \ln D}{\partial \theta}
   -\frac{1}{2}x^T \frac{\partial S}{\partial \theta} x
   +x^T \frac{\partial S}{\partial \theta}\mu
   +x^T S\frac{\partial \mu}{\partial \theta}
   -\frac{\partial \mu^T }{\partial \theta}S \mu
   -\frac{1}{2}\mu^T\frac{\partial S }{\partial \theta}\mu
\end{eqnarray}
we now take the derivative wrt a second parameter $\phi$:

\begin{eqnarray}
\frac{\partial^2 \ln p(x|\theta)}{\partial \theta \partial \phi}
&=&-\frac{1}{2} \frac{\partial^2 \ln D}{\partial \theta \partial \phi}
   -\frac{1}{2}x^T \frac{\partial^2 S}{\partial \theta \partial \phi} x
   +x^T \frac{\partial}{\partial \phi}\left(\frac{\partial S}{\partial \theta}\mu\right)
   +x^T \frac{\partial}{\partial \phi}\left(S\frac{\partial \mu}{\partial \theta}\right)\\
&& -\frac{\partial}{\partial \phi}\left(\frac{\partial \mu^T }{\partial \theta}S \mu\right)
   -\frac{1}{2}\frac{\partial}{\partial \phi}\left(\mu^T\frac{\partial S }{\partial \theta}\mu\right)\\
&=&-\frac{1}{2} \frac{\partial^2 \ln D}{\partial \theta \partial \phi}
   -\frac{1}{2}x^T \frac{\partial^2 S}{\partial \theta \partial \phi} x
   +x^T \left(\frac{\partial^2 S}{\partial \theta \partial \phi}\mu+\frac{\partial S}{\partial \theta}\frac{\partial \mu}{\partial \phi}\right)
   +x^T \left(\frac{\partial S}{\partial \phi}\frac{\partial \mu}{\partial \theta}+S\frac{\partial^2 \mu}{\partial \theta \partial \phi}\right)\\
&&   -\left(\frac{\partial^2 \mu^T }{\partial \theta \partial \phi}S \mu
                    +\frac{\partial \mu^T }{\partial \theta}\frac{\partial S}{\partial \phi} \mu
                    +\frac{\partial \mu^T }{\partial \theta}S \frac{\partial \mu}{\partial \phi}\right)
   -\frac{1}{2}\left(\frac{\partial \mu^T}{\partial \phi}\frac{\partial S }{\partial \theta}\mu
         +\mu^T\frac{\partial^2 S }{\partial \theta \partial \phi}\mu
         +\mu^T\frac{\partial S }{\partial \theta}\frac{\partial \mu}{\partial \phi}\right)\nonumber\\
&=&-\frac{1}{2} \frac{\partial^2 \ln D}{\partial \theta \partial \phi}
   -\frac{1}{2}x^T \frac{\partial^2 S}{\partial \theta \partial \phi} x
   +x^T \frac{\partial^2 S}{\partial \theta \partial \phi}\mu
   +x^T \frac{\partial S}{\partial \theta}\frac{\partial \mu}{\partial \phi}
   +x^T \frac{\partial S}{\partial \phi}\frac{\partial \mu}{\partial \theta}
   +x^T S\frac{\partial^2 \mu}{\partial \theta \partial \phi}\\
&&   -\frac{\partial^2 \mu^T }{\partial \theta \partial \phi}S \mu
     -\frac{\partial \mu^T }{\partial \theta}\frac{\partial S}{\partial \phi} \mu
     -\frac{\partial \mu^T }{\partial \theta}S \frac{\partial \mu}{\partial \phi}
   -\frac{1}{2}\frac{\partial \mu^T}{\partial \phi}\frac{\partial S }{\partial \theta}\mu
   -\frac{1}{2}\mu^T\frac{\partial^2 S }{\partial \theta \partial \phi}\mu
   -\frac{1}{2}\mu^T\frac{\partial S }{\partial \theta}\frac{\partial \mu}{\partial \phi}
\end{eqnarray}

Taking expectations:
\begin{eqnarray}
E\left(\frac{\partial^2 \ln p(x|\theta)}{\partial \theta^2}\right)
&=&-\frac{1}{2} \frac{\partial^2 \ln D}{\partial \theta \partial \phi}
   -\frac{1}{2}E\left(x^T \frac{\partial^2 S}{\partial \theta \partial \phi} x\right)
   +E(x^T) \frac{\partial^2 S}{\partial \theta \partial \phi}\mu
   +E(x^T) \frac{\partial S}{\partial \theta}\frac{\partial \mu}{\partial \phi}\\
&& +E(x^T) \frac{\partial S}{\partial \phi}\frac{\partial \mu}{\partial \theta}
   +E(x^T) S\frac{\partial^2 \mu}{\partial \theta \partial \phi}
     -\frac{\partial^2 \mu^T }{\partial \theta \partial \phi}S \mu
     -\frac{\partial \mu^T }{\partial \theta}\frac{\partial S}{\partial \phi} \mu\\
&&   -\frac{\partial \mu^T }{\partial \theta}S \frac{\partial \mu}{\partial \phi}
   -\frac{1}{2}\frac{\partial \mu^T}{\partial \phi}\frac{\partial S }{\partial \theta}\mu
   -\frac{1}{2}\mu^T\frac{\partial^2 S }{\partial \theta \partial \phi}\mu
   -\frac{1}{2}\mu^T\frac{\partial S }{\partial \theta}\frac{\partial \mu}{\partial \phi}
\end{eqnarray}

But $E(x)=\mu$, and, from appendix 2,
\begin{equation}
E\left(x^T \frac{\partial^2 S}{\partial \theta \partial \phi} x\right)
=\mbox{tr}\left(\frac{\partial^2 S}{\partial \theta \partial \phi}\Sigma \right)+\mu^T \frac{\partial^2 S}{\partial \theta \partial \phi} \mu
\end{equation}

and so:

\begin{eqnarray}
E\left(\frac{\partial^2 \ln p(x|\theta)}{\partial \theta \partial \phi}\right)
&=&-\frac{1}{2} \frac{\partial^2 \ln D}{\partial \theta \partial \phi}
   -\frac{1}{2}\left(\mbox{tr}\left( \frac{\partial^2 S}{\partial \theta \partial \phi}\Sigma\right)+\mu^T \frac{\partial^2 S}{\partial \theta \partial \phi} \mu\right)
   +\mu^T \frac{\partial^2 S}{\partial \theta \partial \phi}\mu
   +\mu^T \frac{\partial S}{\partial \theta}\frac{\partial \mu}{\partial \phi}\\
&& +\mu^T \frac{\partial S}{\partial \phi}\frac{\partial \mu}{\partial \theta}
   +\mu^T S\frac{\partial^2 \mu}{\partial \theta \partial \phi}
     -\frac{\partial^2 \mu^T }{\partial \theta \partial \phi}S \mu
     -\frac{\partial \mu^T }{\partial \theta}\frac{\partial S}{\partial \phi} \mu\\
&&   -\frac{\partial \mu^T }{\partial \theta}S \frac{\partial \mu}{\partial \phi}
   -\frac{1}{2}\frac{\partial \mu^T}{\partial \phi}\frac{\partial S }{\partial \theta}\mu
   -\frac{1}{2}\mu^T\frac{\partial^2 S }{\partial \theta \partial \phi}\mu
   -\frac{1}{2}\mu^T\frac{\partial S }{\partial \theta}\frac{\partial \mu}{\partial \phi}\\
&=&-\frac{1}{2} \frac{\partial^2 \ln D}{\partial \theta \partial \phi}
   -\frac{1}{2}\mbox{tr}\left(\frac{\partial^2 S}{\partial \theta \partial \phi}\Sigma \right)
   -\frac{1}{2}\mu^T \frac{\partial^2 S}{\partial \theta \partial \phi} \mu
   +\mu^T \frac{\partial^2 S}{\partial \theta \partial \phi}\mu
   +\mu^T \frac{\partial S}{\partial \theta}\frac{\partial \mu}{\partial \phi}\\
&& +\mu^T \frac{\partial S}{\partial \phi}\frac{\partial \mu}{\partial \theta}
   +\mu^T S\frac{\partial^2 \mu}{\partial \theta \partial \phi}
     -\frac{\partial^2 \mu^T }{\partial \theta \partial \phi}S \mu
     -\frac{\partial \mu^T }{\partial \theta}\frac{\partial S}{\partial \phi} \mu\\
&&   -\frac{\partial \mu^T }{\partial \theta}S \frac{\partial \mu}{\partial \phi}
   -\frac{1}{2}\frac{\partial \mu^T}{\partial \phi}\frac{\partial S }{\partial \theta}\mu
   -\frac{1}{2}\mu^T\frac{\partial^2 S }{\partial \theta \partial \phi}\mu
   -\frac{1}{2}\mu^T\frac{\partial S }{\partial \theta}\frac{\partial \mu}{\partial \phi}\\
&=&-\frac{1}{2} \left[\frac{\partial^2 \ln D}{\partial \theta \partial \phi}
   +\mbox{tr}\left( \frac{\partial^2 S}{\partial \theta \partial \phi}\Sigma\right)\right]
   -\frac{\partial \mu^T }{\partial \theta}S \frac{\partial \mu}{\partial \phi}\label{eq4}
\end{eqnarray}

But
\begin{eqnarray}
\frac{\partial^2 \ln D}{\partial \theta \partial \phi}
&=&\frac{\partial}{\partial \theta}\left(\frac{\partial \log D}{\partial \phi}\right)\\
&=&\frac{\partial}{\partial \theta}\left(\frac{1}{D}\frac{\partial D}{\partial \phi}\right)\\
&=&\frac{\partial}{\partial \theta}\left(\frac{1}{D}\left(D\mbox{tr}\left(S\frac{\partial \Sigma}{\partial \phi}\right)\right)\right)\\
&=&\frac{\partial}{\partial \theta}\left(\mbox{tr}\left(S\frac{\partial \Sigma}{\partial \phi}\right)\right)\\
&=&\mbox{tr}\left(\frac{\partial}{\partial \theta}\left(S\frac{\partial \Sigma}{\partial \phi}\right)\right)\\
&=&\mbox{tr}\left(\frac{\partial S}{\partial \theta}\frac{\partial \Sigma}{\partial \phi}+S\frac{\partial^2 \Sigma}{\partial \theta \partial \phi}\right)\\
&=&\mbox{tr}\left(\frac{\partial S}{\partial \theta}\frac{\partial \Sigma}{\partial \phi}\right)+\mbox{tr}\left( S \frac{\partial^2 \Sigma}{\partial \theta \partial \phi}\right)
\end{eqnarray}

However
\begin{eqnarray}
\frac{\partial}{\partial \theta}\left(S \Sigma\right)&=&S \frac{\partial \Sigma}{\partial \theta}+\frac{\partial S}{\partial \theta}\Sigma\\
\frac{\partial^2}{\partial \theta \partial \phi}\left( S \Sigma\right)
&=&S \frac{\partial^2 \Sigma}{\partial \theta \partial \phi}
    +\frac{\partial S}{\partial \phi}\frac{\partial \Sigma}{\partial \theta}
    +\frac{\partial^2 S}{\partial \theta \partial \phi}\Sigma
    +\frac{\partial S}{\partial \theta}\frac{\partial \Sigma}{\partial \phi}
\end{eqnarray}

But $S\Sigma=I$ and so $\frac{\partial^2}{\partial \theta \partial \phi}( S \Sigma)=0$, implying that
\begin{equation}
S\frac{\partial^2 \Sigma}{\partial \theta \partial \phi}=- \frac{\partial^2 S}{\partial \theta \partial \phi}\Sigma
-\frac{\partial S}{\partial \theta}\frac{\partial \Sigma}{\partial \phi}
-\frac{\partial S}{\partial \phi}\frac{\partial \Sigma}{\partial \theta}
\end{equation}

and so
\begin{eqnarray}
\frac{\partial^2 \ln D}{\partial \theta \partial \phi}
&=&\mbox{tr}\left(\frac{\partial S}{\partial \theta}\frac{\partial \Sigma}{\partial \phi}\right)+\mbox{tr}\left( S \frac{\partial^2 \Sigma}{\partial \theta \partial \phi}\right)\\
&=&\mbox{tr}\left(\frac{\partial S}{\partial \theta}\frac{\partial \Sigma}{\partial \phi}\right)
+\mbox{tr}\left(- \frac{\partial^2 S}{\partial \theta \partial \phi}\Sigma
 -\frac{\partial S}{\partial \theta}\frac{\partial \Sigma}{\partial \phi}
 -\frac{\partial S}{\partial \phi}\frac{\partial \Sigma}{\partial \theta}\right)\\
&=&\mbox{tr}\left(\frac{\partial S}{\partial \theta}\frac{\partial \Sigma}{\partial \phi}\right)
+\mbox{tr}\left(- \frac{\partial^2 S}{\partial \theta \partial \phi}\Sigma\right)
-\mbox{tr}\left(\frac{\partial S}{\partial \theta}\frac{\partial \Sigma}{\partial \phi}\right)
-\mbox{tr}\left(\frac{\partial S}{\partial \phi}\frac{\partial \Sigma}{\partial \theta}\right)\\
&=&\mbox{tr}\left(- \frac{\partial^2 S}{\partial \theta \partial \phi}\Sigma\right)
-\mbox{tr}\left(\frac{\partial S}{\partial \phi}\frac{\partial \Sigma}{\partial \theta}\right)
\end{eqnarray}

Giving:
\begin{eqnarray}
\frac{\partial^2 \ln D}{\partial \theta \partial \phi}+\mbox{tr}\left(\frac{\partial^2 S}{\partial \theta \partial \phi}\Sigma \right)
&=&\mbox{tr}\left(- \frac{\partial^2 S}{\partial \theta \partial \phi}\Sigma\right)
-\mbox{tr}\left(\frac{\partial S}{\partial \phi}\frac{\partial \Sigma}{\partial \theta}\right)
+\mbox{tr}\left(\frac{\partial^2 S}{\partial \theta \partial \phi}\Sigma \right)\\
&=&-\mbox{tr}\left(\frac{\partial S}{\partial \phi}\frac{\partial \Sigma}{\partial \theta}\right)\label{eq5}
\end{eqnarray}

Substituting expression~\ref{eq5} into equation~\ref{eq4} gives:

\begin{eqnarray}
E\left(\frac{\partial^2 \ln p(x|\theta)}{\partial \theta \partial \phi}\right)
&=&-\frac{1}{2} \left[\frac{\partial^2 \ln D}{\partial \theta \partial \phi}
   +\mbox{tr}\left( \frac{\partial^2 S}{\partial \theta \partial \phi}\Sigma\right)\right]
   -\frac{\partial \mu^T }{\partial \theta}S \frac{\partial \mu}{\partial \phi}\\
&=&\frac{1}{2}\mbox{tr}\left(\frac{\partial S}{\partial \phi}\frac{\partial \Sigma}{\partial \theta}\right)
   -\frac{\partial \mu^T }{\partial \theta}S \frac{\partial \mu}{\partial \phi}\\
&=&-\frac{1}{2}\mbox{tr}\left(S\frac{\partial \Sigma}{\partial \phi}S\frac{\partial \Sigma}{\partial \theta}\right)
   -\frac{\partial \mu^T }{\partial \theta}S \frac{\partial \mu}{\partial \phi}
\end{eqnarray}

We now generalize from two to multiple parameters. We change the notation so that $\theta$ is the vector
of all parameters. The prior is then:
\begin{eqnarray}
p(\theta)
&=&\sqrt{-\mbox{det}E\left(\frac{\partial^2 \ln p(x|\theta)}{\partial \theta_j \partial \theta_k}\right)}\\
&=&\sqrt{\mbox{det}\left(\frac{\partial \mu^T }{\partial \theta_j}S \frac{\partial \mu}{\partial \theta_k}
-\frac{1}{2}\mbox{tr}\left(\frac{\partial S}{\partial \theta_j}\frac{\partial \Sigma}{\partial \theta_k}\right)\right)}\label{eq101}\\
&=&\sqrt{\mbox{det}\left(\frac{\partial \mu^T }{\partial \theta_j}S \frac{\partial \mu}{\partial \theta_k}
+\frac{1}{2}\mbox{tr}\left(S\frac{\partial \Sigma}{\partial \theta_j}S\frac{\partial \Sigma}{\partial \theta_k}\right)\right)}\label{eq101b}
\end{eqnarray}

If there are $n$ observations and $m$ parameters then:
\begin{itemize}
  \item $\mu$ is an $n$ x $1$ vector
  \item $\frac{\partial \mu}{\partial \theta_j}$ is an $n$ by $1$ vector
  \item $\mu^T$ is a $1$ x $n$ vector
  \item $\frac{\partial \mu^T}{\partial \theta_k}$ is a $1$ by $n$ vector
  \item $S$ is an $n$ by $n$ matrix
  \item $\frac{\partial \mu^T }{\partial \theta_j}S \frac{\partial \mu}{\partial \theta_k}$ is a scalar
  (which is the $(j,k)$th element of a matrix)
  \item $\frac{\partial S}{\partial \theta_j}$ is an $n$ by $n$ matrix
  \item $\frac{\partial \Sigma}{\partial \theta_k}$ is an $n$ by $n$ matrix
  \item $\frac{\partial S}{\partial \theta_j}\frac{\partial \Sigma}{\partial \theta_k}$ is an $n$ by $n$ matrix
  \item tr$\left(\frac{\partial S}{\partial \theta_j}\frac{\partial \Sigma}{\partial \theta_k}\right)$ is a scalar
    (which is the $(j,k)$th element of a matrix)
\end{itemize}

We now again consider various special cases of this general formula, starting with the two cases described in~\citet{jp1} and~\citet{jp2}.

\subsection{Independence}

If the observations are modelled as independent then:
\begin{eqnarray}
\Sigma
&=&\mbox{diag}\left(\sigma_1^2,\sigma_2^2,...,\sigma_n^2\right)\\
S
&=&\mbox{diag}\left(\sigma_1^{-2},\sigma_2^{-2},...,\sigma_n^{-2}\right)\\
\frac{\partial \mu^T }{\partial \theta_j}S \frac{\partial \mu}{\partial \theta_k}
&=&\sum_{i=1}^{n}\frac{1}{\sigma_i^2}\left(\frac{\partial \mu_i}{\partial \theta_j}\right)\left(\frac{\partial \mu_i}{\partial \theta_k}\right)\\
\frac{\partial \Sigma}{\partial \theta_k}
&=&\mbox{diag}\left(2\sigma_1 \frac{\partial \sigma_1}{\partial \theta_k},2\sigma_2\frac{\partial \sigma_2}{\partial \theta_k},...,2\sigma_n\frac{\partial \sigma_n}{\partial \theta_k}\right)\\
\frac{\partial S}{\partial \theta_j}
&=&\mbox{diag}\left(-2\sigma_1^{-3} \frac{\partial \sigma_1}{\partial \theta_j},-2\sigma_2^{-3}\frac{\partial \sigma_2}{\partial \theta_j},...,-2\sigma_n^{-3}\frac{\partial \sigma_n}{\partial \theta_j}\right)\\
\frac{\partial S}{\partial \theta_j}\frac{\partial \Sigma}{\partial \theta_k}
&=&\mbox{diag}\left(-4\sigma_1^{-2} \frac{\partial \sigma_1}{\partial \theta_j}\frac{\partial \sigma_1}{\partial \theta_k},-4\sigma_2^{-2}\frac{\partial \sigma_2}{\partial \theta_j}\frac{\partial \sigma_2}{\partial \theta_k},...,-4\sigma_n^{-2}\frac{\partial \sigma_n}{\partial \theta_j}\frac{\partial \sigma_n}{\partial \theta_k}\right)\\
&=&-4\mbox{diag}\left(\frac{1}{\sigma_1^2} \frac{\partial \sigma_1}{\partial \theta_j}\frac{\partial \sigma_1}{\partial \theta_k},\frac{1}{\sigma_2^2}\frac{\partial \sigma_2}{\partial \theta_j}\frac{\partial \sigma_2}{\partial \theta_k},...,\frac{1}{\sigma_n^2}\frac{\partial \sigma_n}{\partial \theta_j}\frac{\partial \sigma_n}{\partial \theta_k}\right)\\
-\frac{1}{2}\mbox{tr}\left(\frac{\partial S}{\partial \theta_j}\frac{\partial \Sigma}{\partial \theta_k}\right)
&=&2\sum_{i=1}^{n}\frac{1}{\sigma_i^2}\frac{\partial \sigma_i}{\partial \theta_j}\frac{\partial \sigma_i}{\partial \theta_k}
\end{eqnarray}

and so
\begin{eqnarray}
p(\theta)&=&\sqrt{\mbox{det}\left(\sum_{i=1}^{n}\frac{1}{\sigma_i^2}\left(\frac{\partial \mu_i}{\partial \theta_j}\right)\left(\frac{\partial \mu_i}{\partial \theta_k}\right)
               +2\sum_{i=1}^{n}\frac{1}{\sigma_i^2}\left(\frac{\partial \sigma_i}{\partial \theta_j}\right)\left(\frac{\partial \sigma_i}{\partial \theta_k}\right)\right)}
\end{eqnarray}

which agrees with equation 36 in~\citet{jp1}.

\subsection{Constant Covariance}

If the covariance $\Sigma$ is constant then equation~\ref{eq101} reduces immediately to
\begin{eqnarray}
p(\theta)&=&\sqrt{\mbox{det}\left(\frac{\partial \mu^T }{\partial \theta_j}S \frac{\partial \mu}{\partial \theta_k}\right)}
\end{eqnarray}

which agrees with equation 28 in~\citet{jp2}.

\subsection{Constant Correlation}

If the correlations are modelled as constant (but the variances are allowed to vary) then:

\begin{eqnarray}
\Sigma&=&VCV\\
\mbox{where }V&=&\mbox{diag}(\sigma_1^2,\sigma_2^2,...,\sigma_n^2)\\
\mbox{and }C&=&\mbox{the correlation matrix}\\
\frac{\partial \Sigma}{\partial \theta_k}
&=& \frac{\partial V}{\partial \theta_k}CV+VC\frac{\partial V}{\partial \theta_k}\\
&=& 2\frac{\partial V}{\partial \theta_k}CV\\
\mbox{where } \frac{\partial V}{\partial \theta_k}
&=&2\mbox{diag}(\sigma_1\frac{\partial \sigma_1}{\partial \theta_k},\sigma_2\frac{\partial \sigma_2}{\partial \theta_k},...,\sigma_n\frac{\partial \sigma_n}{\partial \theta_k})\\
S&=&V^{-1} C^{-1} V^{-1}\\
\mbox{where }V^{-1}&=&\mbox{diag}(\sigma_1^{-2},\sigma_2^{-2},...,\sigma_n^{-2})\\
\mbox{and }C^{-1}&=&\mbox{the inverse correlation matrix}\\
\frac{\partial S}{\partial \theta_j}
&=& \frac{\partial V^{-1}}{\partial \theta_j}C^{-1}V^{-1}+V^{-1}C^{-1}\frac{\partial V^{-1}}{\partial \theta_j}\\
&=& 2V^{-1}C^{-1}\frac{\partial V^{-1}}{\partial \theta_j}\\
\mbox{where } \frac{\partial V^{-1}}{\partial \theta_j}
&=&-2\mbox{diag}\left(\frac{1}{\sigma_1^2}\frac{\partial \sigma_1}{\partial \theta_j},\frac{1}{\sigma_2^2}\frac{\partial \sigma_2}{\partial \theta_j},...,\frac{1}{\sigma_n^2}\frac{\partial \sigma_n}{\partial \theta_j}\right)\\
\frac{\partial V^{-1}}{\partial \theta_j}\frac{\partial V}{\partial \theta_k}
&=&-4\mbox{diag}\left(\frac{1}{\sigma_1}\frac{\partial \sigma_1}{\partial \theta_j}\frac{\partial \sigma_1}{\partial \theta_k},\frac{1}{\sigma_2}\frac{\partial \sigma_2}{\partial \theta_j}\frac{\partial \sigma_2}{\partial \theta_k},...,\frac{1}{\sigma_n}\frac{\partial \sigma_n}{\partial \theta_j}\frac{\partial \sigma_n}{\partial \theta_k}\right)\\
\frac{\partial S}{\partial \theta_j}\frac{\partial \Sigma}{\partial \theta_k}
&=&4 V^{-1}C^{-1}\frac{\partial V^{-1}}{\partial \theta_j}\frac{\partial V}{\partial \theta_k}CV
\end{eqnarray}

and the prior is:
\begin{eqnarray}
p(\theta)
&=&\sqrt{\mbox{det}\left(\frac{\partial \mu^T }{\partial \theta_j}S \frac{\partial \mu}{\partial \theta_k}
-\frac{1}{2}\mbox{tr}\left(4 V^{-1}C^{-1}\frac{\partial V^{-1}}{\partial \theta_j}\frac{\partial V}{\partial \theta_k}CV\right)\right)}
\end{eqnarray}

\subsection{Constant Variance}

If the variances are modelled as constant (but the correlations are allowed to vary) then:
\begin{eqnarray}
\Sigma&=&VCV\\
\mbox{where }V&=&\mbox{diag}(\sigma_1^2,\sigma_2^2,...,\sigma_n^2)\\
\mbox{and }C&=&\mbox{the correlation matrix}\\
\frac{\partial \Sigma}{\partial \theta_k}&=& V \frac{\partial C}{\partial \theta}V\\
S&=&V^{-1} C^{-1} V^{-1}\\
\mbox{where }V&=&\mbox{diag}(\sigma_1^{-2},\sigma_2^{-2},...,\sigma_n^{-2})\\
\mbox{and }C{-1}&=&\mbox{the inverse correlation matrix}\\
\frac{\partial S}{\partial \theta_j}&=&V^{-1} \frac{\partial C^{-1}}{\partial \theta_j}V^{-1}\\
\frac{\partial S}{\partial \theta_j}\frac{\partial \Sigma}{\partial \theta_k}
&=&V^{-1}\frac{\partial C^{-1}}{\partial \theta_j}VV^{-1}\frac{\partial C}{\partial \theta_k}V\\
&=&V^{-1}\frac{\partial C^{-1}}{\partial \theta_j}\frac{\partial C}{\partial \theta_k}V\\
&=&-V^{-1}C^{-1}\frac{\partial C}{\partial \theta_j}C^{-1}\frac{\partial C}{\partial \theta_k}V
\end{eqnarray}

and the prior is:
\begin{eqnarray}
p(y|x)
&=&\sqrt{\mbox{det}\left(\frac{\partial \mu^T }{\partial \theta_j}S \frac{\partial \mu}{\partial \theta_k}
-\frac{1}{2}\mbox{tr}\left(V^{-1}\frac{\partial C^{-1}}{\partial \theta_j}\frac{\partial C}{\partial \theta_k}V\right)\right)}\\
&=&\sqrt{\mbox{det}\left(\frac{\partial \mu^T }{\partial \theta_j}S \frac{\partial \mu}{\partial \theta_k}
+\frac{1}{2}\mbox{tr}\left(V^{-1}C^{-1}\frac{\partial C}{\partial \theta_j}C^{-1}\frac{\partial C}{\partial \theta_k}V\right)\right)}
\end{eqnarray}

\section{Summary}\label{s5}

Climate models are statistical models, in that they produce probabilistic predictions
(when run as initial condition ensembles) and have certain parameters that can only
be determined by comparison of model results with observations.
As a result, the Bayesian framework, in which probabilistic predictions made from models with different parameter values are
combined together to make a single best probabilistic prediction, can be applied.
Within that framework one has to specify a prior, and one must choose between a prior based on intuition or a prior
based on a rule. The former is known as subjective Bayesian statistics, and the latter, objective Bayesian statistics.
The authors are pursuing a research programme that is exploring methods by which objective Bayesian statistics can be applied in
climate modelling. In this article we have discussed the application of the most standard objective prior, known as
Jeffreys' Prior.

Climate models are complex and the relationship between the parameters
and the predicted distributions is also complex. However, the form of the predicted distributions themselves can often
be rather simple, and for many groups of variables a multivariate normal may be a good approximation.
We have shown that, by making this approximation, the calculation of Jeffreys' Prior can be reduced to
differentiating the parameters of the multivariate normal by the parameters of the underlying climate model.
We derive expressions for Jeffreys' Prior in this situation.

The results from our two previous articles on this topic (\citet{jp1} and~\citet{jp2}) are special cases of the general results shown here.

In all this work we have expressed Jeffreys' Prior in terms of the true, rather than the estimated, parameters of the distributions from climate model predictions. In other words, we have assumed infinite rather than finite size initial condition ensembles.
A further challenge is to rederive the expressions given above but incorporating estimation uncertainty.

\appendix

\section{Proof that if $S$ is symmetric then $a^TSb=b^TSa$}

Since $a^TSb$ is a scalar it is equal to its tranpose, as so:
\begin{equation}
a^TSb=(a^TSb)^T=b^TS^Ta
\end{equation}
but if $S$ is symmetric then
\begin{equation}
b^TS^Ta=b^TSa
\end{equation}
Putting these two together gives:
\begin{equation}
a^TSb=b^TSa
\end{equation}

\section{Proof that $E(x^TAx)=\mbox{tr}\left(\Sigma A\right)+\mu^TA\mu$}

\begin{eqnarray}
E(x^TAx)
&=&E\left(\sum_{ij} x_i A_{ij} x_j\right)\\
&=&E\left(\sum_{ij} x_i x_j A_{ij} \right)\\
&=&\sum_{ij} E(x_i x_j) A_{ij}
\end{eqnarray}

But, by definition:
\begin{eqnarray}
\Sigma_{ij}
&=&E((x_i-\mu_i)(x_j-\mu_j))\\
&=&E(x_i x_j-x_i \mu_j-\mu_i x_j+\mu_i \mu_j)\\
&=&E(x_i x_j)-E(x_i) \mu_j-\mu_i E(x_j)+\mu_i \mu_j)\\
&=&E(x_i x_j)-\mu_i \mu_j-\mu_i \mu_j+\mu_i \mu_j)\\
&=&E(x_i x_j)-\mu_i \mu_j
\end{eqnarray}
So
\begin{eqnarray}
E(x_i x_j)&=&\Sigma_{ij}+\mu_i \mu_j
\end{eqnarray}

and so
\begin{eqnarray}
E(x^TAx)
&=&\sum_{ij} (\Sigma_{ij}+\mu_i \mu_j) A_{ij}\\
&=&\sum_{ij} \Sigma_{ij}A_{ij}+\sum_{ij}\mu_i\mu_j a_{ij}\\
&=&\mbox{tr}\left(\Sigma A\right)+\mu^T A \mu\\
&=&\mbox{tr}\left(A\Sigma \right)+\mu^T A \mu
\end{eqnarray}

\bibliography{arxiv}

%\newpage
%\input{AJOTablesV7}

\end{document}